\documentclass{article}
\usepackage[utf8]{inputenc}
\usepackage[a4paper, total={6in, 8in}]{geometry}
\usepackage{amsmath,amssymb,amsfonts}
\usepackage{placeins}
\usepackage{authblk}
\usepackage{graphicx,color}
\usepackage{xcolor}
\usepackage{listings}
\usepackage{siunitx}
\usepackage{hhline}
\usepackage{hyperref}
\usepackage{wrapfig}
\usepackage{cite}
\usepackage{algorithmic}
\usepackage{textcomp}

\begin{document}

\title{Merging in a Coupled Driving Simulator: \\ How do drivers resolve conflicts?}

\author[1]{Olger Siebinga\thanks{Corresponding author}}
\author[1]{Arkady Zgonnikov}
\author[1]{David A. Abbink}
\date{July 2023}

\affil[1]{Human-Robot Interaction Group, Department of Cognitive Robotics, Faculty of Mechanical Engineering, Delft University of Technology, Delft, the Netherlandss}
\renewcommand\Affilfont{\itshape\small}

\maketitle

\begin{abstract}
Traffic interactions between merging and highway vehicles are a major topic of research, yielding many empirical studies and models of driver behaviour. Most of these studies on merging use naturalistic data. Although this provides insight into human gap acceptance and traffic flow effects, it obscures the operational inputs of interacting drivers. Besides that, researchers have no control over the vehicle kinematics (i.e., positions and velocities) at the start of the interactions. Therefore the relationship between initial kinematics and the outcome of the interaction is difficult to investigate. To address these gaps, we conducted an experiment in a coupled driving simulator with a simplified, top-down view, merging scenario with two vehicles. We found that kinematics can explain the outcome (i.e., which driver merges first) and the duration of the merging conflict. Furthermore, our results show that drivers use key decision moments combined with constant acceleration inputs (intermittent piecewise-constant control) during merging. This indicates that they do not continuously optimize their expected utility. Therefore, these results advocate the development of interaction models based on intermittent piecewise-constant control. We hope our work can contribute to this development and to the fundamental knowledge of interactive driver behaviour. \\
\end{abstract}

\section{Introduction}
Interactions between vehicles, such as in highway merging, play a major role in everyday traffic. Therefore, driving behaviour in these interactions is an essential aspect of many transportation technologies. Empirical data and microscopic traffic models of human driving behaviour are thus essential tools for transportation engineers. These models and data are used in the design and safety assessment of highway on-ramps~\cite{Hassan2012, Lord2005} and urban intersections~\cite{Wang2018}. Microscopic traffic models can be used to evaluate traffic management systems~\cite{Yang1996}. And finally, autonomous vehicle designers are interested in these interactions to develop socially acceptable and human-like autonomous behaviour~\cite{Schwarting2019, Sadigh2018}. Particularly for the last use case, a good understanding of the individual negotiations and the continuous reciprocal actions of the drivers during interactions is essential.

Many recent studies have investigated interactive merging behaviour by modelling this behaviour or by conducting empirical investigations. Most of these studies use naturalistic data, i.e., data recorded in real-world scenarios. For example, Daamen et al.~\cite{Daamen2010} and Marczak et al.~\cite{Marczak2013} performed empirical analysis on traffic data which they recorded with helicopters. Wang et al.~\cite{Wang2022} and Srinivasan et al.~\cite{Srinivasan2021} used existing open datasets to evaluate driver behaviour on merge ramps. Others have modelled interactive driver behaviour using naturalistic data to gain insights, e.g., using game theory~\cite{Ji2020a, Liu2007, Kang2017}, acceleration models comparable to car-following models~\cite{wan2014}, or machine-learned models~\cite{Dong2018, Srinivasan2021}.

\begin{figure*}[ht!]
\centering
\begin{minipage}{.65\textwidth}
    \centering
    \includegraphics[width=\linewidth]{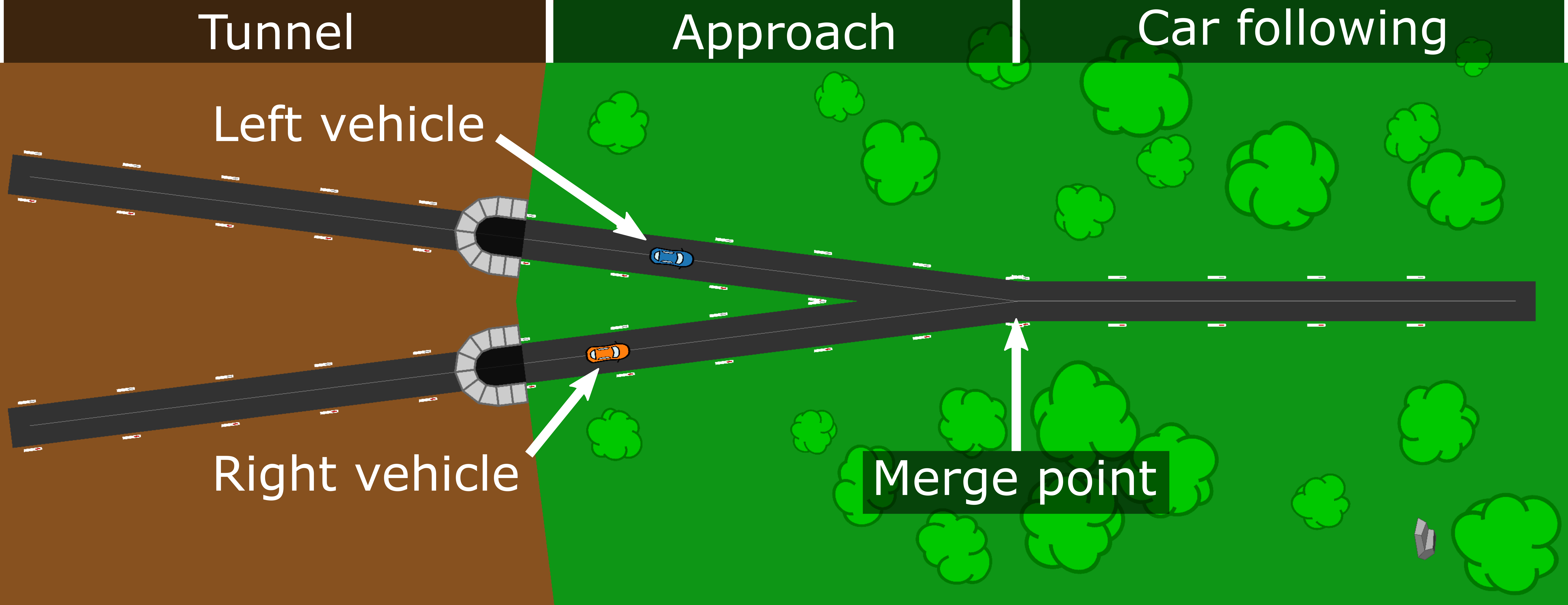}
    \caption{The simplified merging scenario used in the experiment. Two vehicles approach a pre-defined merge point at which their lanes merge into one. The track consists of three sections of equal length ($50~m$, total track length $150~m$). The vehicle dimensions are $4.5~m$ x $1.8~m$. In the tunnel, participants could observe both vehicles, but not control their vehicles. During the approach, the participants could control the acceleration of their vehicles to resolve the merging conflict. During the car-following section, the vehicles follow each other in the same lane.}
    \label{fig:track_overview}
\end{minipage}
\hfill
\begin{minipage}{.25\textwidth}
    \centering
    \includegraphics[width=0.6\linewidth]{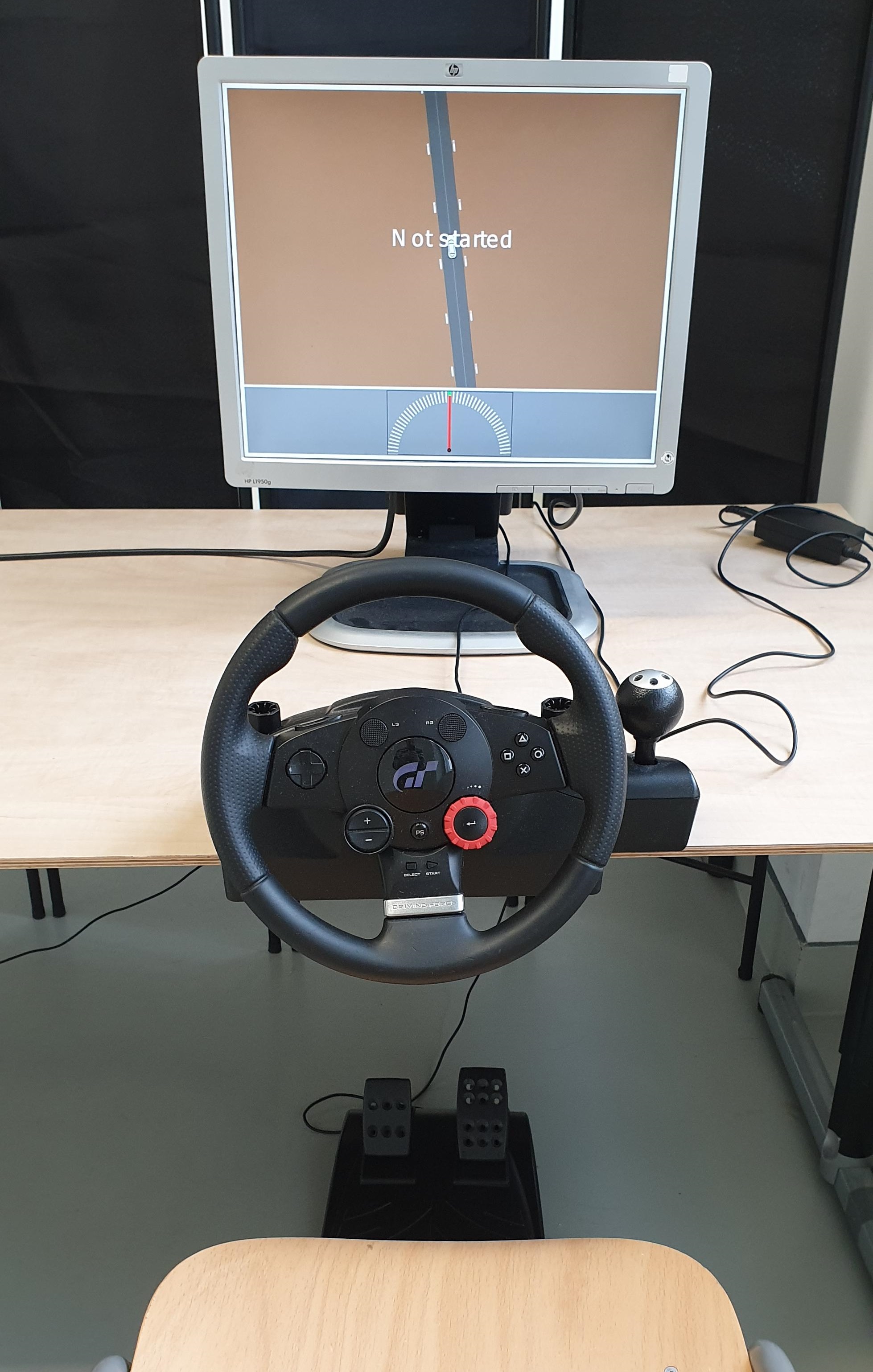}
    \caption{The experimental setup as seen from a participant's view. The other participant in the pair used an identical setup. The participants could not see each other.}
    \label{fig:setup}
\end{minipage}
\end{figure*}

The usage of naturalistic data has the advantage that real-world behaviour can be studied. However, this approach has two main drawbacks. First, naturalistic data is recorded with cameras on helicopters, quad-copters, or high buildings. Therefore, only sequential positions are recorded. Velocities and accelerations are reconstructed from this position data. This makes it challenging to directly investigate the drivers' operational behaviour and control inputs. Second, kinematic differences between conditions can be observed, but not controlled. This makes it difficult to investigate the relationship between the initial kinematics of the vehicles and the outcome of the merging conflict (e.g., who merges first and who yields). To gain a deeper understanding of individual reciprocal interactions, controlled experiments are needed.

However, only a very limited number of studies in a controlled environment (i.e., in a driving simulator) targeted interactions during merging (i.e., excluding studies of autonomous control strategies, gap acceptance, or traffic flow). Stoll et al. investigated human decision-making in merging scenarios based on videos of a controlled simulation~\cite{Stoll2020}. Participants had to select their preferred reaction (e.g., accelerate or decelerate) after watching videos of vehicles they were "interacting with". Shimojo et al. used a driving simulator to investigate how the merging behaviour of drivers is affected by their perception of other drivers~\cite{Shimojo2022}. They used predetermined controls for one of the vehicles in the interaction, to influence this perception in a controlled way. In both experiments, the behaviour of one of the drivers was predetermined. Thus, there was no interaction or dynamic negotiation between two human drivers. We conclude that the existing literature misses studies that investigate the reciprocal merging interactions between at least two human drivers in a controlled environment.

To address this gap, we conduct an experiment in a top-down view, coupled driving simulator in which we investigate reciprocal merging interactions between two human drivers. We investigate the operational behaviour of the drivers in terms of inputs (acceleration and velocity profiles). Furthermore, we examine the influence of different initial kinematics (both position and velocity) on the outcome of the interaction. Both on a high level in terms of which driver merges first, and in more detail through the metric Conflict Resolution Time (CRT)~\cite{Siebinga2022d}. We hope this experiment advances the fundamental knowledge about vehicle-vehicle interactions in traffic and contributes to the development of interaction-aware intelligent transportation systems.

\section{Methods}
We conducted an experiment in a coupled, top-down view driving simulator with 9 pairs of participants (6 female, 12 male, mean age: 25, std: 2.6). The details of this experiment (including Figures~\ref{fig:track_overview} and~\ref{fig:setup}), and the analysis tools we developed to gain insight into the merging behaviour, have been previously published in~\cite{Siebinga2022d}. This experiment was approved by TU Delft's Human Research Ethics Committee (HREC). All participants gave their consent before participating in the experiment.

The experiment regarded a symmetric simplified merging scenario (Figure~\ref{fig:track_overview}) in which participants could control the acceleration of their vehicle using the gas and brake pedal of a steering-wheel game controller (Logitech Driving Force GT). The headings of the vehicles were always equal to the heading of the road, so no steering was involved. Participants could see the simulation on a computer screen (Figure~\ref{fig:setup}). However, they could not see the other participant, who was seated in the same room behind a screen. To prevent auditory communication, participants wore noise-cancelling headsets (Sony WH-1000XM3) with ambient music. All gathered data was published in the 4TU data repository~\cite{Siebinga2022a}. The software needed to reproduce the experiment can be found on GitHub\footnote[1]{\url{https://github.com/tud-hri/simple-merging-experiment}}. Interactive plots of all our results can be found in the online supplementary materials\footnote[2]{\url{https://tud-hri.github.io/simple-merging-experiment}}.

To investigate the effects of the initial vehicle kinematics on the outcome of the merging conflict we varied the initial positions and initial velocities of the vehicles. Participants were instructed to maintain their initial velocity yet prevent a collision. To ensure a merging conflict, all conditions were chosen such that if both drivers would maintain their initial velocity, they would collide. Furthermore, participants were instructed to remain seated, use one foot on the gas or brake pedal, keep both hands on the steering wheel, and not to communicate by making sounds or noise. Finally, participants were told that this is a scientific experiment --not a game or a race-- and that no vehicle had the right of way.

The participants received visual feedback on their computer screens. Their visuals were randomly mirrored such that they appeared to approach the merge point from the left or the right side randomly. While on the experimenter's view, and in all results discussed here, we refer to the same participant in a pair as the left or right driver. If participants deviated from their initial velocity, their steering wheel provided vibration feedback, increasing with the deviation and with a dead band around the initial velocity. If the vehicles collided, the participants got a time penalty of 20 seconds. This was longer than the duration of a single trial, which took approximately 16 seconds.

The vehicles started in a tunnel where participants could observe the initial velocities of both vehicles. But they could not control their vehicle yet. Once both vehicles had exited the tunnel, both participants gained control. This marked an unambiguous moment when the interaction started. The vehicles' initial positions and velocities were varied to create 11 experimental conditions. We used the projected headway at the merge point as the underlying metric to design the conditions and determine the initial positions. The projected headway is the headway (distance from front bumper to front bumper) at the merge point if both drivers would maintain their initial velocity. We chose this metric because it does not depend on track dimensions or a snapshot of the vehicle state at an arbitrary point along the track (e.g., at the tunnel exit).

\begin{figure}[h]
    \centering
    \includegraphics[width=0.6\linewidth]{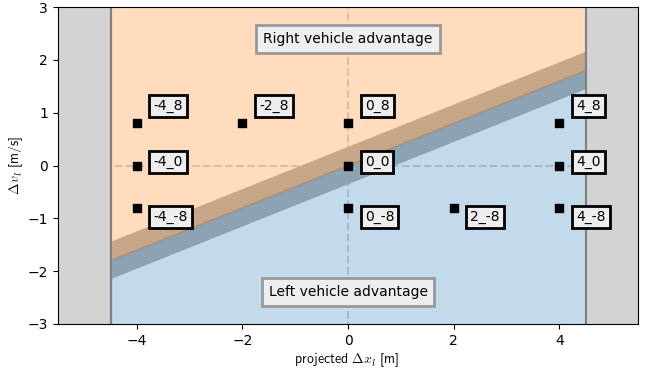}
    \caption{The experimental conditions in their two-dimensional space. The x-axis shows the projected headway at the merge point if both drivers would keep their initial velocity. If the headway is larger than the vehicle length ($4.5~m$) there is no projected collision, this is indicated by the grey areas on the left and right side. The y-axis shows the initial velocity differences. Positive values mean that the left vehicle is (projected to be) ahead or moving faster. The diagonal darker area divides the space into areas where the left or right driver has the advantage to pass the merge point first. This line was estimated based on pilot experiments. Note that this does not simply divide the plane into areas where one driver has the velocity or projected headway advantage.}
    \label{fig:conditions}
\end{figure}

To visualise the differences between conditions, we plotted them in a 2D projected-headway - relative-velocity plane (Figure~\ref{fig:conditions}). This figure shows the conflict space. If the projected headway is larger than the vehicle length, there is no conflict. These areas are shown in grey on the left and right side of Figure~\ref{fig:conditions}. The figure also shows in which areas we expected the right or the left driver to have an advantage. This expectation was based on a (shorter) pilot experiment with the same experimental setup but different kinematic conditions. 

We used this expectation to design and spread the conditions evenly over the conflict space. The diagonal darker area represents the area in which the (kinematic) advantage changes from the left to the right driver. We decided not to investigate this area but to (first) focus on driver behaviour in cases where the outcome is more distinct. Our aim here is to gain insight into the interactions and negotiations between the two drivers in these situations. However, we did include a baseline condition where neither driver has a position or velocity advantage. With these conditions, we aim to obtain a quantitative description of the most likely outcome (who merges first) based on the initial kinematics. We used the Python package Pymer4~\cite{Jolly2018} for all statistical models in this work.

\begin{figure}[ht]
    \centering
    \includegraphics[width=.9\linewidth]{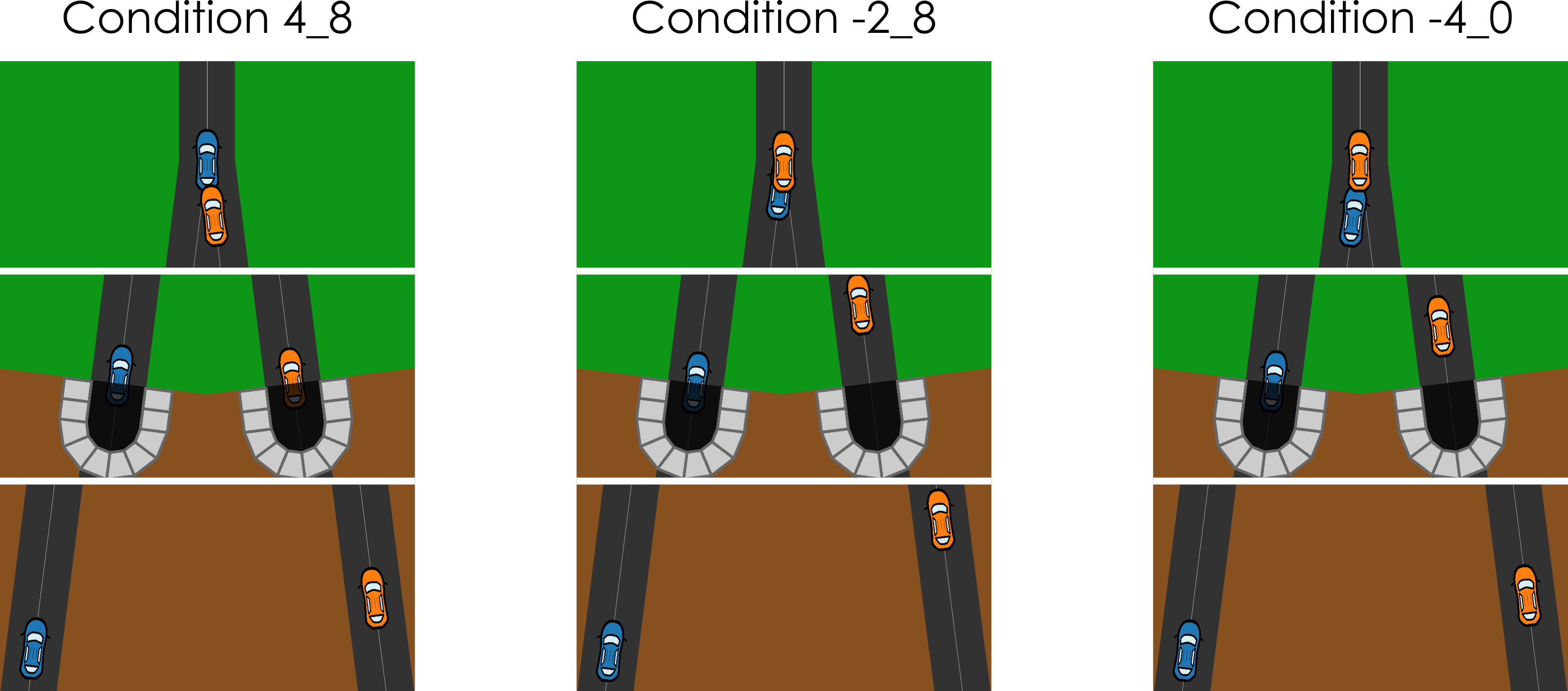}
    \caption{Three visualisations of experimental conditions. The figures show the relative positions of the vehicles and the start point, tunnel exit, and merge point. These merge point positions would occur if both vehicles would maintain their initial velocity. In most conditions, the slower vehicle has a position advantage at the tunnel exit. The exceptions are conditions $4\_8$ and $-4\_-8$, where the vehicles exit the tunnel at the same time.}
    \label{fig:condition_examples}
\end{figure}

We named the conditions based on the two dimensions that define them: the projected headway in meters, and the velocity difference in decimeters per second. Positive numbers indicate that the left driver has an advantage. For example, in condition \textbf{-2\_8}, the right driver has a projected headway advantage of $2~m$, but the left driver drives $0.8~\frac{m}{s}$ faster. For more visual examples of conditions and their names, see Figure~\ref{fig:condition_examples}. In our experiment, every condition was repeated 10 times in a random order for every pair of participants.


We used the Conflict Resolution Time (CRT)~\cite{Siebinga2022d} to analyse the conflict resolution behaviour of the pairs of participants. The CRT denotes the time from the start of the interaction until the first moment at which the vehicles are no longer on a collision course (assuming constant velocity). To calculate the CRT, we post-process the data and determine for every time step if a collision would occur on the remaining track if both vehicles would continue their velocity. The time between the tunnel exit and the first moment where no collision would occur is the CRT. Thus, CRT is a measure of the amount of time needed to resolve the conflict and therefore can be used as a measure of the difficulty of the merging conflict. 

\section{Results}
\begin{table*}[ht!]
    \caption{The number of observed collisions per condition. The total number of trials per condition was 90. Most collisions occurred between 4 and 6 seconds after the vehicles exited the tunnels.}
    \label{tab:collisions}
    \centering
    \begin{tabular}{|c|c|c|c|c|c|c|c|c|c|c|c|} \hline
         Condition &  -4\_-8 & -4\_0 & -4\_8 & -2\_8 & 0\_8 & 0\_0 & 0\_-8 & 2\_-8 & 4\_-8 & 4\_0 & 4\_8\\ \hline
         Number of Collisions & 3 & 1 & 3 & 2 & 4 & 5 & 3 & 2 & 1 & 2 & 2 \\ \hline
    \end{tabular}
\end{table*}

We structure our investigation of driver conflict resolution behaviour into two parts. First, we present the analysis of the joint behaviour of two drivers, to analyse the outcome of the conflict (who gives way) and how quickly each pair of drivers resolved the merging conflict. Metrics that capture the joint behaviour for each pair under different conditions include a percentage of who merged first, as well as the Conflict Resolution Time (CRT). Second, we investigate the contributions of each individual driver in a pair to resolve the conflict. This includes the actions the individual drivers took in terms of accelerations and the resulting velocity profiles. 

\subsection{Joint Behaviour}
\subsubsection{Who Merged First?}

\begin{figure}[h!]
    \centering
    \includegraphics[width=0.6\linewidth]{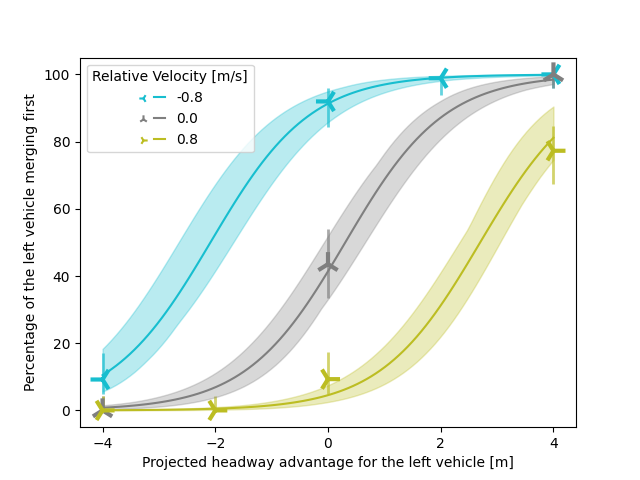}
    \caption{An overview of the high-level outcome per condition: which driver went first? Every condition was repeated 10 times for all 9 participant pairs. Therefore, the total number of trials per condition is 90. The markers show the measured data as the percentage of the left driver merging first, with the vertical line representing the $95\%$ binomial proportion confidence intervals. Collisions were omitted from these results (see Table~\ref{tab:collisions}). The lines and shaded areas represent the (population) predictions of the mixed-effects logistic regression model (Table~\ref{tab:first_regression}) with the $95\%$ confidence interval.}
    \label{fig:who_first}
\end{figure}

The high-level outcome of a merging conflict can be summarised by which driver reached the merge point first, except for the trials where the vehicles collided. However, collisions were rare across all conditions (Table~\ref{tab:collisions}). We plot the proportion of left and right vehicles that went first as a function of initial conditions in Figure~\ref{fig:who_first}. In the "neutral" $0\_0$ condition this proportion is almost evenly distributed. For the other 10 conditions with kinematic differences between the drivers, 5 conditions show a consistent outcome over all pairs and trials. This indicates that the outcome in these conditions is entirely defined by kinematics, with no variation between participant pairs. In one other condition ($2\_-8$), only a single trial deviated from the outcome norm. Four conditions ($-4\_-8$, $4\_8$, $0\_-8$, and $0\_8$) show a large majority of the outcomes where a particular driver merges first and a minority of the other driver merging first.  

\begin{table*}[]
\centering
\caption{Mixed-effects logistic regression model describing the effect of projected headway and relative velocity on which driver reached the merge point first. Collisions were excluded, the left vehicle going first was labelled as 1, right first as 0. The model includes a random intercept for participant pairs to account for between-pair differences.}
\label{tab:first_regression}
\begin{tabular}{l|l|l|l|l|ll|}
                  &         &         &         &         & \multicolumn{2}{l|}{Confidence interval} \\
                  & Estimate    & SE & Z       & P-value & \multicolumn{1}{l|}{0.025}    & 0.975    \\ \hhline{=|=|=|=|=|=|=|}
Intercept         & -0.32 & 0.212   & -1.50  & $\num{1.326e-01}$   & \multicolumn{1}{l|}{-0.73}   & 0.10   \\
Projected headway & 1.15  & 0.080  & 14.4  & $\num{6.966e-47}$   & \multicolumn{1}{l|}{0.99}    & 1.31    \\
Relative velocity & -3.4138 & 0.321   & -10.6 & $\num{2.858e-26}$   & \multicolumn{1}{l|}{-4.04}   & -2.78   \\ \hline
\end{tabular}
\end{table*}

\begin{table*}[]
\centering
\caption{Fixed effects estimates of the random intercept values per pair for the mixed-effects logistic regression model (Table~\ref{tab:first_regression}).}
\label{tab:individual_intercepts}
\begin{tabular}{|l|c|c|c|c|c|c|c|c|c|} \hline
Participant Pair & 1 & 2 & 3 & 4 & 5 & 6 & 7 & 8 & 9 \\ \hline
Intercept & $-0.54$ & $-0.42$ & $-1.17$ & $0.06$ & $-0.13$ & $-0.51$ & $0.16$ & $-0.22$ & $-0.13$ \\ \hline
\end{tabular}
\end{table*}

\begin{figure*}[h!]
    \centering
    \includegraphics[width=0.9\textwidth]{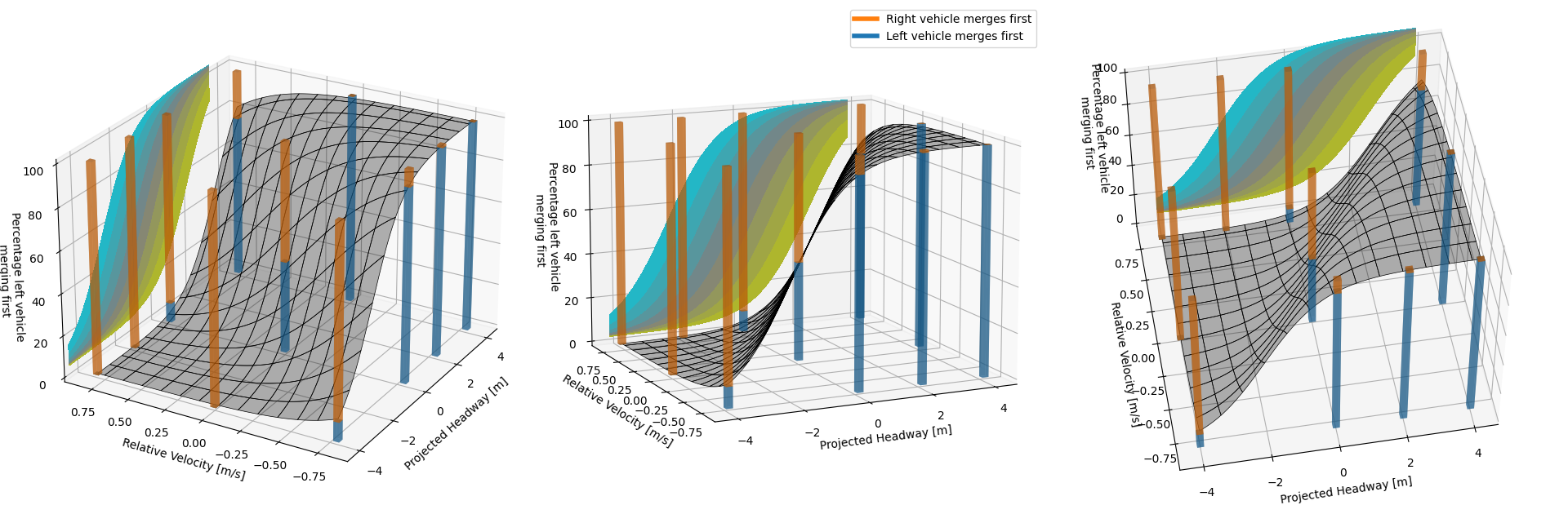}
    \caption{A 3-dimensional visualisation of a (population) prediction of the logistic regression model on the data. All three subplots show the same data for different angles. The model predictions are shown as the black surface and the background projections. The coloured bars show the data from the experiment. The $x$ and $y$-axis represent the condition kinematics. The $z$-axis shows the percentage of trials where the left driver merged first. Collisions were excluded from this data (see Table~\ref{tab:collisions}). An interactive version of this plot can be found in the online supplementary materials.}
    \label{fig:surface_plot}
\end{figure*}

To investigate the relationship between the initial conditions (i.e. the kinematics at the start of each scenario) and the outcome (which driver merges first), we fitted a mixed-effects logistic regression model to the data. The model parameters are shown in Table~\ref{tab:first_regression}, and the model outcome is visualised in Figures~\ref{fig:who_first} and~\ref{fig:surface_plot}. These results show that increasing the projected headway advantage increases the chances of a driver merging first ($z=14.4,\,p<10^{-46}$). The relative velocity on the other hand has a negative effect on the probability the driver merges first ($z=-10.6,\,p<10^{-25}$). This means that, for equal projected headways, a driver with a higher initial velocity tends to merge behind the driver with a lower initial velocity. The explanation for this is that drivers with a  higher velocity exit the tunnel later than the slower vehicle in most conditions (Figure~\ref{fig:condition_examples}). An important side-note to these effects is that we found these in a symmetric scenario with no right of way for either of the drivers.

The population level intercept had a negative estimated value that is not significant ($z=-1.5,\, p=0.13$). This could be explained by the fact that the intercept explains a bias in the data towards the left or the right driver. This effect is clearest in the neutral condition ($0\_0$), where we found that the right driver merged first in a small majority of the cases. Table~\ref{tab:individual_intercepts} shows the estimated intercept values for the individual participant pairs. We expect that with more participants, the bias on the population level will disappear and the intercept value will approach 0.

\begin{figure}[h]
    \centering
    \includegraphics[width=0.7\linewidth]{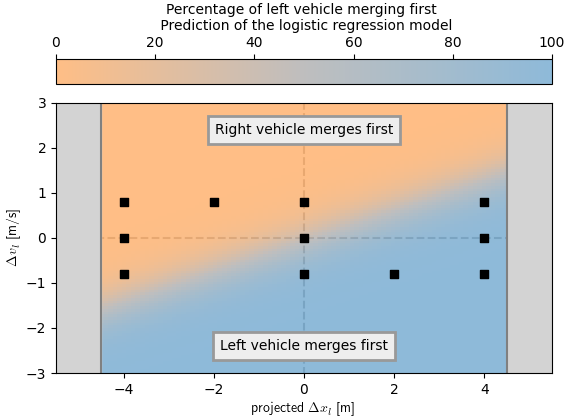}
    \caption{A heat map of a logistic regression model prediction for the driver that will merge first. The conditions where data was gathered are marked with black squares.}
    \label{fig:heatmap}
\end{figure}

To visualise at which locations in the conflict space the left or right driver is more likely to merge first, we have created a top-down view heat map of the regression model. This heat map is shown in Figure~\ref{fig:heatmap} and closely resembles Figure~\ref{fig:conditions}.

\subsubsection{Conflict Resolution Time}
\begin{figure*}[h!]
    \centering
    \includegraphics[width=\textwidth]{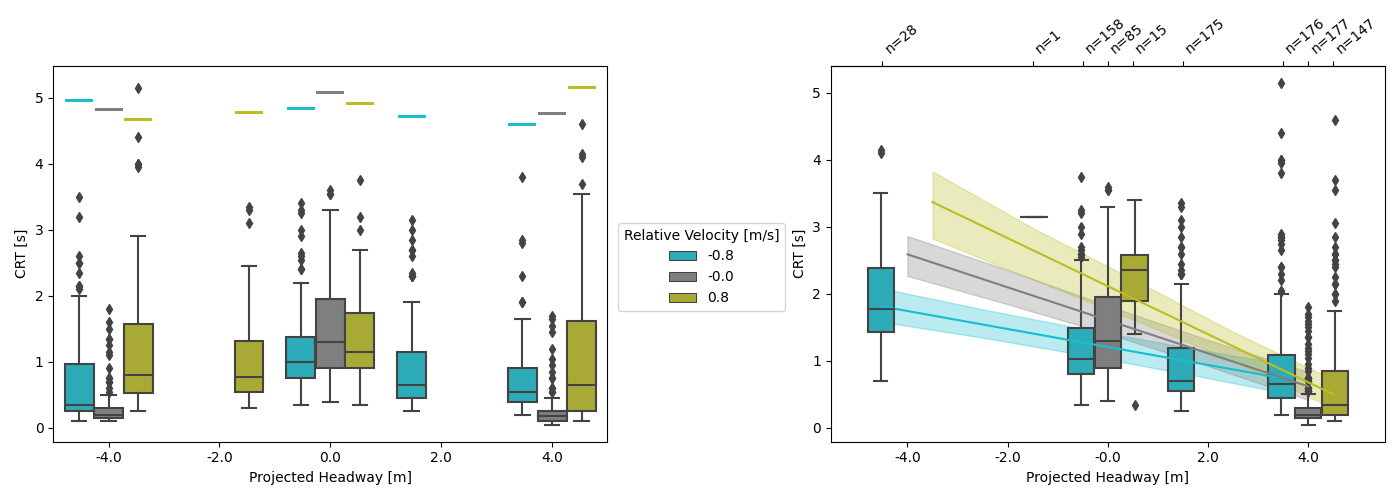}
\hspace{0.03\textwidth}
\begin{minipage}{0.45\textwidth}
    \centering
    \caption[Distribution of the Conflict Resolution Time (CRT) for all conditions]{Distribution of the Conflict Resolution Time (CRT) for all conditions. The CRT is the time from the moment at which the drivers gain control until the first moment when they are no longer on a collision course (assuming constant velocities). The coloured horizontal bars indicate the average time at which the first vehicle reached the merge point in that condition. A figure that shows the same CRT distribution placed in the 2-dimension conflict space on the locations of the corresponding conditions is available in the online supplementary material.}
    \label{fig:crt}
\end{minipage}
\hspace{0.12\textwidth}
\begin{minipage}{0.4\textwidth}
    \centering
    \caption{Distribution of the Conflict Resolution Time (CRT) from the perspective of the first merging driver. In this plot, positive numbers for headway and velocity differences indicate an advantage for the driver that merged first in that trial. This results in a different number of trials per box (see the labels at the top of the figure). The lines and shaded areas visualise predictions of the mixed effects model (Table~\ref{tab:crt_model}) and its 95\% confidence interval.}
    \label{fig:crt_model}
\end{minipage}
\end{figure*}

Besides how the conflict was resolved (which driver merged first) we investigated how quickly the conflict was resolved by examining the Conflict Resolution Time (CRT). This is a measure of the time it took the drivers to resolve the conflict and therefore resembles the difficulty of the conflict in a specific trial. Figure~\ref{fig:crt} shows the CRT distributions we found for all experimental conditions. The median CRT is highest for the neutral condition $0\_0$. In this condition, no driver has a headway or velocity advantage. Drivers have to negotiate a solution without a "most-likely" candidate solution. The lowest median CRT was found for the conditions where one driver only had a projected headway difference but the velocity was the same for both drivers. The conditions with velocities differences but no projected headway difference had high median CRTs. Thus conflicts where one driver has a pure projected headway advantage are easier to resolve than conflicts where one driver has a pure velocity advantage.

But besides these high-level observations, Figure~\ref{fig:crt} reveals no clear relationship between the initial kinematics and the CRT of the merging conflicts. We expected that the high-level outcome of the conflict (who merged first) might partly explain the CRT of that trial. More concretely, we expected trials where the driver with the kinematic advantage went first, to be resolved more quickly than trials where the driver with a disadvantage went first. To investigate this, we analysed CRT as a function of the kinematic advantage from the perspective of the first merging driver (Figure~\ref{fig:crt_model}, Table~\ref{tab:crt_model}). The projected headway and velocity differences in this figure are positive if the first merging driver had the advantage. We found that trials with a larger headway advantage for the driver that merged first had a lower CRT ($t=-15.3,\,p<10^{-46}$). Trials with a velocity advantage for the first merging driver had a higher CRT ($t=5.02,\,p<10^{-6}$). Moreover, we found that the association between the CRT and the projected headway advantage was stronger for larger velocity advantage ($t=-6.09,\,p<10^{-8}$). One important side note is that drivers with a higher initial velocity have a headway disadvantage in the approach section, i.e., they are approaching the merge point behind the other driver. 

\begin{table*}[]
\centering
\caption{Mixed-effects linear regression model analysing the Conflict Resolution Time (CRT) as a function of the kinematic conditions. Positive headways and relative velocities indicate an advantage for the driver who merged first. Collisions were excluded.}

\label{tab:crt_model}
\begin{tabular}{|p{1.8in}|l|l|l|l|ll|} \hline
                  &         &         &         &         & \multicolumn{2}{l|}{Confidence interval} \\ 
                  & Estimate    & SE & T-stat       & P-value & \multicolumn{1}{l|}{0.025}    & 0.975    \\ \hhline{=|=|=|=|=|=|=|}
Intercept         & 1.61 & 0.107   & 15.2  & $\num{4.97e-10}$   & \multicolumn{1}{l|}{1.41}   & 1.83   \\ \hline
Projected headway & -0.25  & 0.016  & -15.3  & $\num{2.16e-47}$   & \multicolumn{1}{l|}{-0.28}    & -0.22    \\ \hline
Relative velocity & 0.40 & 0.080   & 5.02 & $\num{6.07e-07}$   & \multicolumn{1}{l|}{0.25}   & 0.56   \\ \hline
Relative velocity : projected headway & -0.14 & 0.023   & -6.09 & $\num{1.68e-9}$   & \multicolumn{1}{l|}{-0.18}   &  -0.09  \\ \hline
\end{tabular}
\end{table*}

\subsection{Individual behaviour}

\begin{figure*}[h!]
    \centering
    \includegraphics[width=0.9\textwidth]{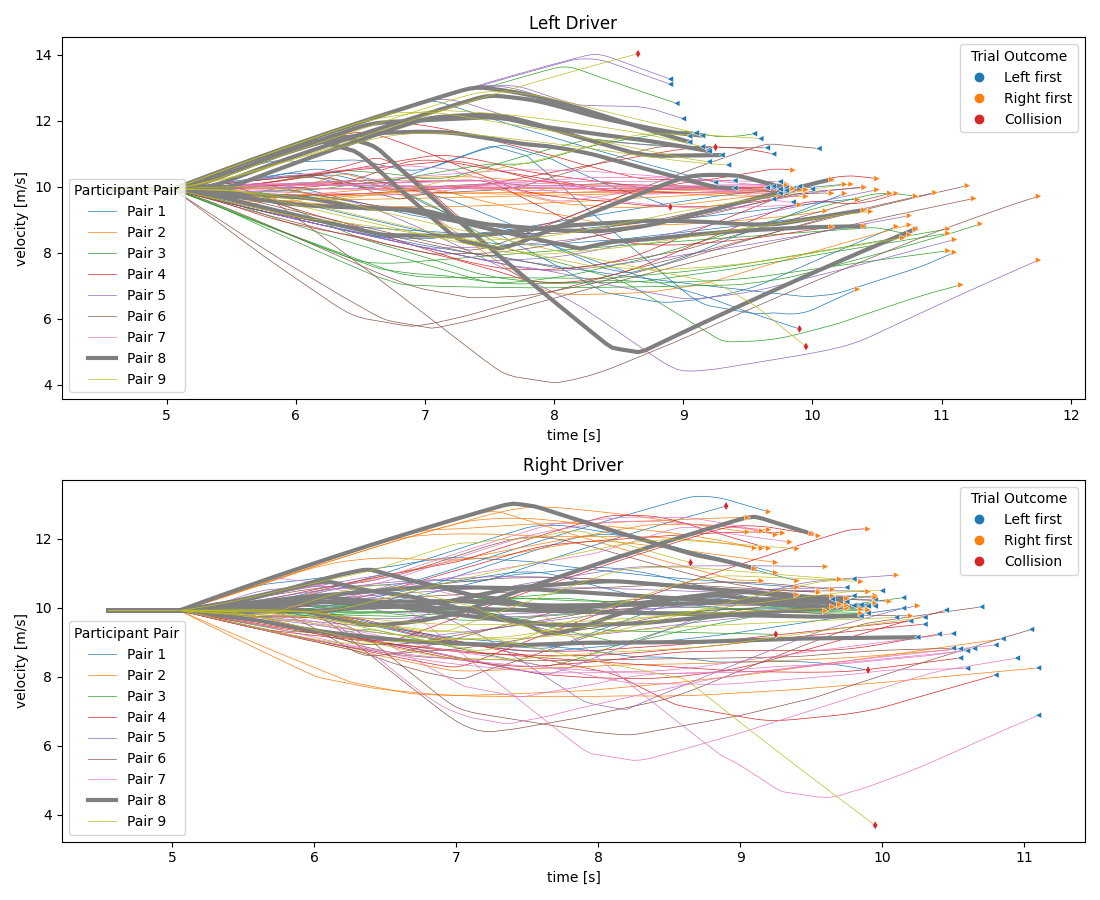}
    \caption{Velocity traces of the left and right drivers for all trials in the neutral condition 0\_0, from the tunnel exit up until the merge point. The trials of a representative pair are highlighted to provide more insight into individual traces. The markers at the end of the trials indicate the final outcome of the trial. These plots show that drivers use triangular velocity patterns while interacting. These triangular patterns indicate that drivers use blocks of constant acceleration input with key decision moments in between. Interactive versions of these plots for all conditions are available in the online supplementary material.}
    \label{fig:velocities}
    \includegraphics[width=\textwidth]{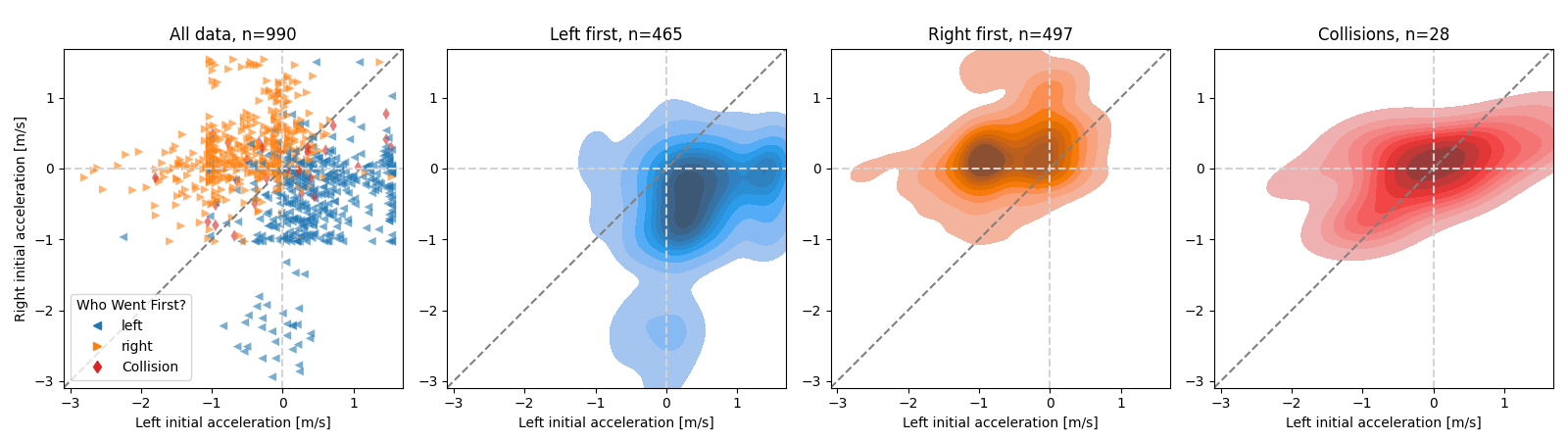}
    \caption{The outcome of the merging conflict plotted versus the initial acceleration input at tunnel exit for the left (x-axis) and right (y-axis) drivers for all conditions.}
    \label{fig:initial_action}
\end{figure*}

To gain insight into the operational behaviour of the drivers, we investigated the aggregated velocity traces of all drivers (Figure~\ref{fig:velocities}). We choose to show the velocity traces for the neutral condition ($0\_0$) here because this condition has the widest variety of solutions (in terms of who merges first). Because of this spread, this velocity plot is easier to read than the same plot for other conditions. However, the key aspects identified in this plot are representative of the other conditions (for the raw data, including plots, see~\cite{Siebinga2022a}. Interactive versions of these plots are available in the supplementary material).

One of the striking characteristics of the velocity traces in Figure~\ref{fig:velocities} are the triangular patterns that can be observed in many traces. Such triangular-shaped velocity patterns indicate two things. First, it shows that drivers use blocks of constant acceleration (step inputs on gas/brake) to control their vehicle during an interaction. Second, in between these step inputs, or straight lines in the velocity trace, the input changes rapidly, causing a sharp angle in the velocity trace. This indicates that drivers select an input level and stick to that until something triggers a new decision resulting in a new input level. We refer to this combination as \textit{intermittent piecewise-constant control}, where intermittent refers to the observed decision moments, and piecewise-constant to the constant acceleration levels in between.

With this intermittent piecewise-constant control, drivers use key decision moments at which they determine a plan. After this decision, they stick with this plan until something triggers a new decision. Therefore, Figure~\ref{fig:velocities} provides evidence that drivers do not continuously optimise their acceleration input while interacting in traffic. Thus, the assumption of continuous utility maximisation that is made in many models of driver behaviour (e.g.~\cite{Siebinga2022, Ji2020, Schwarting2018, Sadigh2018, Naumann2020}) does not hold for these interactions. 

Another aspect shown in Figure~\ref{fig:velocities} is that in many cases, the drivers immediately accelerate or decelerate at the moment they gain control. This indicates, that even in this purely symmetrical condition, drivers exit the tunnel with an intended solution in mind (i.e., they plan to go first or yield). To further investigate if drivers start the interaction with a mutual solution in mind, and if this solution is also reached, we plotted the outcome of the merging conflict versus the initial drivers' actions in Figure~\ref{fig:initial_action}.

Figure~\ref{fig:initial_action} shows that in the majority of the interactions that do not end in a collision, the drivers initially cooperate. In most interactions that end in the left vehicle reaching the merge point first, the left driver's initial input was to accelerate and the right driver's initial input was to decelerate. This indicates two things. First, it shows that drivers form compatible ideas about who will merge first before they even start interacting (in that trial), i.e., drivers use a shared mental model~\cite{Jonker2010}. Second, even though there are cases where the conflict is resolved by only one of the drivers (i.e., where the other driver's input is 0), in most cases, both drivers initially act simultaneously to prevent a collision. 

\FloatBarrier
\section{Discussion}
In this paper, we investigated the conflict-resolving behaviour of pairs of drivers in a simplified merging scenario. Our four most important findings are: 1) both the relative velocity and projected headway have a significant effect on which driver merges first; 2) the time it takes drivers to resolve the conflict (CRT) can be explained by the kinematics from the perspective of the driver that merges first; 3) drivers used a shared mental model about which driver merges first based on observations before the start of the interaction; and 4) drivers use intermittent piecewise-constant control to resolve the conflict. suggesting they do not constantly optimise some utility function. Rather the observed control behaviour is in line with satisficing (see ~\cite{Simon1956}): in our experiment drivers seem to search for a plan that is good enough and stick to that plan until it no longer suffices. At this key decision moment, they re-plan to find a new input that is good enough, and act accordingly.

\subsection{Relation to the Existing Literature}
Our study indicated for the first time that both the relative velocity and the projected headway significantly influence which driver merges first. A velocity advantage decreases the probability of a vehicle merging first while a projected headway advantage increases that probability. Earlier studies mostly used naturalistic data, where these kinematics can not be controlled (e.g.,~\cite{Daamen2010, Marczak2013, Klitzke2022, Wang2022}), or reduced the analysis of kinematics to one dimension by studying time to arrival (e.g.,~\cite{Srinivasan2021}).

The finding that humans do not constantly optimise their behaviour corresponds to previous findings in simple economic games~\cite{Camerer2003}, velocity choice for isolated drivers~\cite{Schmidt-Daffy2014}, and high-level skill switching (between manual braking and using cruise control) during driving~\cite{Goodrich2000}. The key-decision moments with constant inputs in-between have previously been observed in individual truck driver behaviour~\cite{Markkula2014}. However, our results are the first to show that these operational aspects of human driving are also present in merging interactions in a controlled experiment. 

Previous empirical studies on merging behaviour used naturalistic data~\cite{Daamen2010, Marczak2013, Srinivasan2021, Klitzke2022, Wang2022}, in which these operational aspects are not included. Most of these studies focus on evaluating gap acceptance behaviour and were inspired by an interest in the effects of merging behaviour on traffic flow~\cite{Daamen2010, Marczak2013, Klitzke2022}. Among the existing studies of naturalistic merging conflicts, two --in particular-- had a goal similar to ours: to understand the dynamics of drivers' conflict-resolving behaviour.

Wang et al.~\cite{Wang2022} studied social interactions on congested highways in the INTERACTION dataset~\cite{Zhan2019}. They divided the merges based on the social preference of the drivers of trough-lane vehicles. Drivers that overtook a merging car before the merge, were labelled "rude", while drivers that slowed down to let the merging vehicle in were labelled "courteous". However, our results indicate that the outcome (who goes first) for most scenarios we tested depends strongly on the kinematic vehicle states \textit{at the start} of the interaction, not on individual differences between drivers. Because the kinematics are not controlled for in naturalistic data, these form a substantially confounding factor which merits caution when attributing driving style as key factor for merging outcomes.

Srinivasan et al.~\cite{Srinivasan2021} used naturalistic data to evaluate a machine-learned model of human merging behaviour. They concluded that this machine-learned model can successfully predict the trajectories shown by drivers in scenarios where one of the vehicles has a large kinematic advantage. Compared to our work, they reduced the kinematic differences to a single dimension: time-to-arrival. A $0.0~s$ time-to-arrival difference corresponds to a $0.0~m$ projected headway in our work, but other time-to-arrival differences can be obtained with multiple combinations of projected headway and relative velocity. Our results show that these both have a significant impact on the outcome of the conflict in terms of the driver that merges first (Table~\ref{tab:first_regression}) and on the CRT (Table~\ref{tab:crt_model}). An important difference between our work and~\cite{Srinivasan2021} is that we only regarded situations where the drivers are on a collision course from the start of the interaction while~\cite{Srinivasan2021} regards large(r) kinematic differences. Nevertheless, we advocate using both relative velocity and projected headway for the kinematic analysis, because they have different effects on the outcome of the interaction. Besides that, we expect no major implications for machine-learned models of human behaviour based on our results.

\subsection{Implications}
However, when regarding approaches that are not purely data-driven, our results could have major implications for models and control strategies. Many driver models make the assumption that humans behave as rational utility maximisers (e.g.,~\cite{Naumann2020, Kita1999, Liu2007, Siebinga2022}). And because these models make this assumption, many control strategies for autonomous vehicles in mixed traffic were proposed that make the same assumption (e.g.,~\cite{Sadigh2018, Schwarting2019, Fisac2019, Coskun2019, Garzon2019, Isele2019}). 

Roughly, two kinds of rational utility maximization are used in driver models. First, there are the models that regard merging as a single high-level decision about who merges first, such as Kita already proposed in 1999~\cite{Kita1999}. Second, there are models that assume drivers continuously optimize some reward function to determine their current input (Naumann et al. showed many examples of reward functions used for this approach in 2020~\cite{Naumann2020}). Our results have major implications for both assumptions.

For models that regard merging as a single decision, our exploration of different kinematic conditions provides valuable insights into driver behaviour. Our results confirm that the vehicles' kinematics at the start of the interaction have a major impact on which driver merges first. This is in line with the model proposed by Kita~\cite{Kita1999}. However, our results also show that the individual differences in outcomes between pairs of drivers are restricted to a limited range of kinematic scenarios. In most scenarios, the same driver merges first for all driver pairs. This would indicate that modelling the decision of who merges first based on individual preferences (differences in reward function) is only valuable for a limited set of conditions where the kinematic differences are small.

For models that assume continuous optimisation, our results have more far-reaching implications. The aggregated velocity plot (Figure~\ref{fig:velocities}) shows that drivers do not continuously optimise, but re-plan at specific decision moments. This indicates that the assumptions that drivers \textbf{continuously} either: optimize, approximately optimize (up to a threshold), or noisily optimise their inputs are not consistent with driver behaviour. Instead, drivers seem to be triggered to change their behaviour at a certain point (at which they might partially optimise to find a new plan). Besides the key-decision moments, Figure~\ref{fig:velocities} also shows piecewise-linear velocity patterns. This indicates that the assumption that drivers aim to minimise a squared difference between their current and desired velocity (as used in many models, e.g.,~\cite{Naumann2020, Siebinga2021}) is also inconsistent with driver behaviour because that would lead to non-linear velocity profiles. 

In general, our findings imply that the mathematical convenience related to main assumptions in game-theoretic models comes at a serious cost to their descriptive power. Thus, although game-theoretic approaches can be very valuable to determine optimal control decisions between rational agents (e.g., in vehicle-to-vehicle communication approaches~\cite{Talebpour2015,Elhenawy2015a,Banjanovic-Mehmedovic2016}), we advise caution in applying them to predicting driver behaviour (either in driver models or in AV control).

\subsection{Recommendations, Limitations, and Future Work}
Therefore, we interpret our results as an encouragement to develop new types of traffic interaction models that do allow for intermittent piecewise-constant control in operational behaviour. Siebinga et al. previously proposed a model framework that could describe intermittent control in traffic interactions~\cite{Siebinga2023}. But there are other (existing) lines of research that also hold potential for application to interactive scenarios, such as evidence accumulation models (e.g.,~\cite{Markkula2023, Zgonnikov2022, Durrani2021}). Besides the intermittent control, new interaction models should use piecewise-constant acceleration as control inputs. Furthermore, they should be able to describe the most likely outcomes for different initial kinematics, independent of individual driver differences (Figure~\ref{fig:who_first}).

Although our work might provide inspiration for the development of novel interaction models, it also has some limitations. The main limitation is the simplification of the merging scenario. We started with a simplified symmetric merging scenario to investigate operational behaviour independent of factors such as a right-of-way. In real-world merging, this does play an important role in the interaction. Furthermore, we reduced the control inputs to acceleration and deceleration only (no steering). This was done to simplify the analysis of the experiment. This design choice decreased the difficulty of the task which could have increased performance. Finally, we used a top-down-view simulation for simplicity. This makes it easier for participants to estimate velocities, which could have decreased the variability in the outcomes. So although this experiment provided valuable insights, more work is needed to validate these results in a more realistic scenario. This validation should be done in a coupled driver simulator where the experimenter has full control over the initial kinematics to make a useful comparison. 

Besides the simplification of the scenario, another limitation is that participants in this experiment knowingly executed 110 merging manoeuvres against the same opponent. This could have influenced the outcome because the participants could have learned the other driver's behaviour. An experiment with more than two drivers and an experimental setup with random pairing could account for this. 

\section{Conclusion}
In this paper, we investigated how drivers resolved merging conflicts in a coupled, top-down view driving simulator. We used a simplified merging scenario that only includes longitudinal control. We investigated driver behaviour under initial conditions with varying relative velocities and projected headways. We used mixed effects regression models, the concept of Conflict Resolution Time (CRT), and aggregated velocity plots to gain insight into driver behaviour. For the experimental conditions studied, we conclude:

\begin{itemize}
    \item Drivers used intermittent control (modifying acceleration only at key decision moments) to resolve merging conflicts. This suggests that drivers do not behave as continuous rational utility maximisers in merging interactions.
    \item Drivers use piecewise-constant acceleration control (blocks of continuous acceleration) resulting in triangular velocity patterns to control their vehicle.
    \item Relative velocity and projected headway are good predictors of which driver is most likely to merge first. They have different effects and are thus both needed for a reliable prediction (instead of reducing the kinematics to a single time-to-arrival value).
    \item We used a metric to describe the amount of time the drivers need to resolve a merging conflict (CRT). We found CRT is associated with the outcome of the interaction combined with the initial kinematic differences (projected headway and relative velocity).
    \item Conditions where one driver has a pure projected headway advantage are resolved faster than conditions with a pure velocity advantage. 
    \item Drivers used shared mental models and observations before the start of the interaction to determine which driver will merge first.
\end{itemize}

\section*{Acknowledgements}
We thank Alexis Derumigny for the advice on statistical modelling. This research was funded by Nissan Motor Company, Ltd. and by the RVO grant
TKI2012P01.

\FloatBarrier
\bibliographystyle{ieeetr}
\bibliography{My_Collection}

\end{document}